\documentclass[10pt,twocolumn,letterpaper]{article}

\usepackage{iccv}              % To produce the CAMERA-READY version
\usepackage{graphicx}
\usepackage{amsmath}
\usepackage{amssymb}
\usepackage{booktabs}
\usepackage{xcolor}
\usepackage{url}

\usepackage{stfloats}

\definecolor{iccvblue}{rgb}{0.21,0.49,0.74}
\usepackage[pagebackref,breaklinks,colorlinks,allcolors=iccvblue]{hyperref}

\usepackage[capitalize]{cleveref}
\crefname{section}{Sec.}{Secs.}
\Crefname{section}{Section}{Sections}
\Crefname{table}{Table}{Tables}
\crefname{table}{Tab.}{Tabs.}

\def\paperID{*****} % *** Enter the Paper ID here
\def\confName{ICCV}
\def\confYear{2025}

\begin{document}

%%%%%%%%% TITLE - PLEASE UPDATE
\title{Global-to-Local or Local-to-Global? Enhancing Image Retrieval with Efficient Local Search and Effective Global Re-ranking}

\author{
Dror Aiger$^1$ \qquad Bingyi Cao$^2$ \qquad Kaifeng Chen$^2$ \qquad Andre Araujo$^2$ \\
$^1$Google \qquad $^2$Google DeepMind \\
{\tt\small \{aigerd, bingyi, francischen, andrearaujo\}@google.com}
}
\maketitle

%%%%%%%%% ABSTRACT
\begin{abstract}

% Our work explores how local and global features can be combined to improve image retrieval performance.
% Traditionally, local features have been computationally expensive, limiting their use to post-processing re-ranking. However, recent advances in efficient local feature algorithms have opened up new possibilities. Building on this, we propose a novel method that merges initial local feature rankings with state-of-the-art global features using multidimensional scaling (MDS). This combined approach has shown significant improvements in benchmarks like the Oxford and Paris datasets. Furthermore, we introduce a new retrieval system based on landmark MDS. This system efficiently indexes and queries images using any black-box pairwise similarity, even learned models. Our experiments with AMES~\cite{suma2024ames} learned pairwise similarity demonstrate the potential of this approach. In a way, our new retrieval system can be viewed as a way to compute global features, for both index and query images, based on in-hand similarities (in contrast to a generic learning of embedding). We believe this research has the potential to significantly advance image retrieval capabilities. 

The dominant paradigm in image retrieval systems today is to search large databases using global image features, and re-rank those initial results with local image feature matching techniques.
This design, dubbed \emph{global-to-local}, stems from the computational cost of local matching approaches, which can only be afforded for a small number of retrieved images.
However, emerging efficient local feature search approaches have opened up new possibilities, in particular enabling detailed retrieval at large scale, to find partial matches which are often missed by global feature search.
In parallel, global feature-based re-ranking has shown promising results with high computational efficiency.
In this work, we leverage these building blocks to introduce a \emph{local-to-global} retrieval paradigm, where efficient local feature search meets effective global feature re-ranking.
Critically, we propose a re-ranking method where global features are computed on-the-fly, based on the local feature retrieval similarities.
Such re-ranking-only global features leverage multidimensional scaling techniques to create embeddings which respect the local similarities obtained during search, enabling a significant re-ranking boost.
Experimentally, we demonstrate solid retrieval performance, setting new state-of-the-art results on the Revisited Oxford and Paris datasets.

\end{abstract}
\section{Introduction}
\label{sec:intro}

Searching vast image databases efficiently with a query picture enables a number of multimodal applications, e.g.~visual shopping~\cite{liu2016cvpr,peng2021rp2k,song2016deep}, fine-grained entity identification~\cite{ypsilantis2023uned,ypsilantis2021met,krause20133d}, knowledge-based visual question answering~\cite{chen2023infoseek,hu2023reveal,mensink2023encvqa}, among others.
Today, such image retrieval systems are generally designed leveraging a \textit{global-to-local} paradigm~\cite{suma2024ames,lee2022cvnet,tan2021rrt,cao2020delg}, where \textit{global} image features are used in a first search stage and \textit{local} image features are used to re-rank the initial list of retrieved candidates via detailed matching.
This approach benefits from the discriminative power and compactness of global representations for large-scale similarity computation, combined with the localized similarity verification capabilities of local representations.

\begin{figure}[t]
    \centering
    \includegraphics[width=8.3cm]{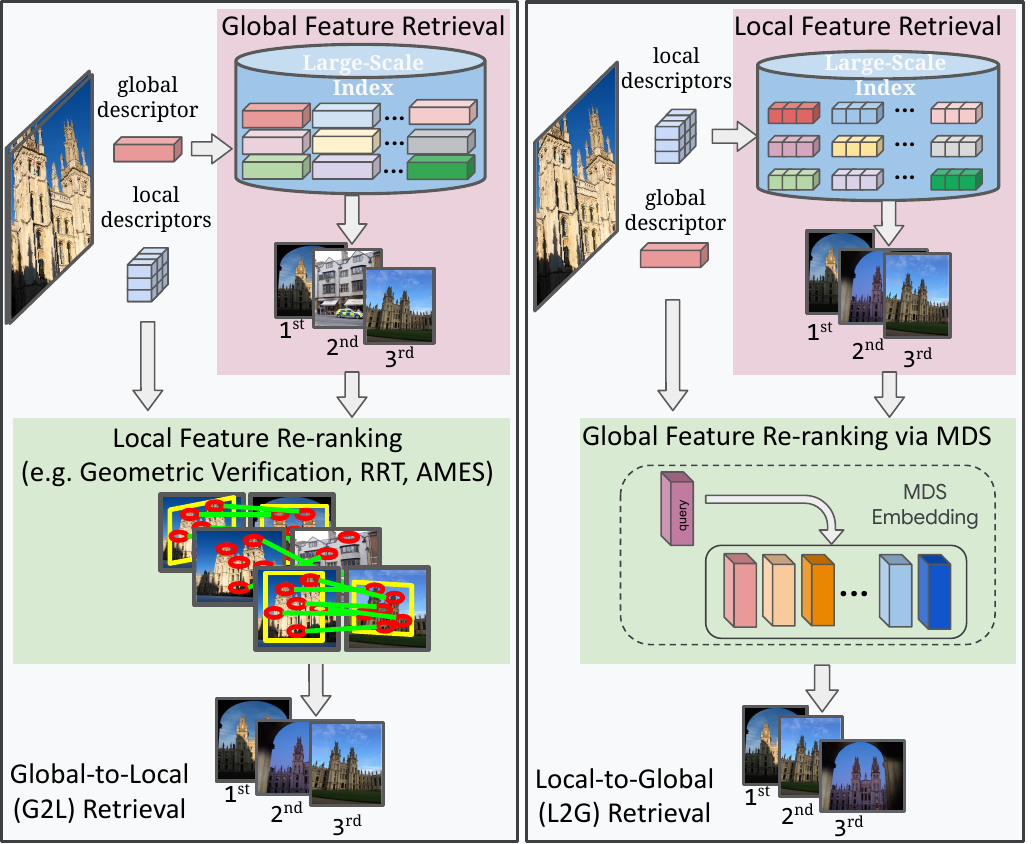}
    \caption{In contrast to conventional Global-to-Local (G2L) image retrieval systems (left), where global features are used for the initial search followed by re-ranking with local features, we introduce a new \textbf{Local-to-Global (L2G)} paradigm (right). 
    In L2G, efficient retrieval with local features meets effective re-ranking with global features.
    Critically, we propose a \textbf{novel global feature re-ranking stage} leveraging multidimensional scaling (MDS) to create query-specific re-ranking embeddings, which are sensitive to localized similarities.
    Experimentally, our system improves upon the state of the art significantly.
    % , and then the retrieved candidates are reranked by a newly created global feature, e.g. MDS embedding in the setup proposed by our paper. %\andre{TODO: Bingyi and/or Kaifeng to draw teaser figure.}
    }
    \label{fig:teaser}
    \vspace{-10pt}
\end{figure}

Despite the success of this framework, a significant concern is the lack of localized search capabilities at large scale, which lead to recall losses at the global feature search stage.
For example, when the query image only has partial matches with relevant database images, the system is usually unable to return pertinent results due to the limitations of global feature similarity estimation.
Besides, while the local feature-based re-ranking stage helps refining the initial matches, it generally does not leverage information across the shortlisted database images to enhance the refinement process, which ends up limiting the final performance.

% \andre{TODO: add a reference to the teaser figure somewhere}

In this work, we address these challenges by proposing a \textit{local-to-global} (L2G) retrieval system, illustrated in Fig. \ref{fig:teaser}.
For the initial search stage, we build on top of recent advances in scalable \textit{local} feature retrieval~\cite{aiger2023cann}, which enables localized search at large scale, enhancing the initial set of retrieved candidates.
We then develop a novel \textit{global} feature re-ranking process, which allows information sharing between the shortlisted images, based on detailed local similarities.
More precisely, we propose to leverage multidimensional scaling (MDS)~\cite{nasir2018mds} to create a global feature for re-ranking purposes, on-the-fly, and to feed it into an effective re-ranking process~\cite{shao2023superglobal}.
MDS enables us to compute global features which respect the detailed local feature similarities.
%The key advantage of our proposed system is that it can efficiently retrieve and re-rank based on localized and detailed similarities, where in particular the re-ranking process can refine relevance scores with a customized embedding that approximates rich local similarities between a given query and its most similar database images.
In summary, the key advantage of our proposed system is that it can efficiently retrieve and re-rank based on localized and detailed similarities -- in particular, the re-ranking process can refine relevance scores with a customized embedding that approximates rich local similarities between a given query and its most relevant database images.
%perform large-scale search using localized representations, then re-rank the shortlisted images by refining their relevance scores with an embedding that approximates rich local similarities. 

\noindent\textbf{Contributions.}
This paper makes three main contributions:

\textbf{(1)}
We propose a new \textit{local-to-global} (L2G) image retrieval paradigm, which flips the conventional script, leveraging local feature search and global feature re-ranking.
Our system enables image retrieval based on localized and precise similarities, which is generally difficult to achieve with previous methods.

\textbf{(2)}
A critical component to make the L2G paradigm work effectively is a new re-ranking method, which creates global features on-the-fly at query time, respecting local similarities.
Leveraging multidimensional scaling, this process creates re-ranking embeddings specific to a given query, allowing us to jointly process the shortlisted images and reorder them effectively.

\textbf{(3)} 
We showcase strong performance on the conventional Revisited Oxford and Paris datasets~\cite{radenovic2018revisiting}, with $2-3$\% %\andre{final numbers} 
gain with respect to previous work, setting a new state of the art.

\section{Related Work} \label{sec:rw}

\noindent\textbf{Image retrieval and re-ranking systems} have a long history in computer vision, even before deep learning techniques dominated the field.
Initial promising results in this area were dominated by \textit{local-to-local} methods, where local feature search was used to find candidates and re-ranking employed geometric verification \cite{Lowe2004,Philbin07,Nister2006vocabulary}.
While these traditionally leveraged hand-crafted local features~\cite{Lowe2004,bay2008speeded}, they were later revisited in the deep learning era with deep image features~\cite{noh2017large,aiger2023cann}.
Little by little, direct local search techniques gave way to methods which aggregated hand-crafted local features into a global feature for search~\cite{Sivic2003,jegou2012aggregating,tolias2015image}, instantiating the first \textit{global-to-local} systems.
These had the advantages of much simpler and lighter search mechanisms, also delivering improved recall.
Such aggregation techniques were also shown effective with deep learned features~\cite{teichmann2019detect,tolias2020learning,weinzaepfel2022learning}.
Most of the deep learning work for image retrieval, though, has been focused on enhancing global features~\cite{babenko2014neural,gordo2017end,Revaud2019ICCV,Ng2020SOLARSL,yang2021dolg}, which generally surpassed the performance of local aggregation techniques.
This also led to a deep learned version of the \textit{global-to-local} paradigm, with both global and local features being extracted in the same model~\cite{cao2020delg,lee2022cvnet}, possibly with additional learnable modules for local similarity estimation~\cite{suma2024ames,tan2021rrt}.
More recently, researchers demonstrated that global features can also be used effectively for the re-ranking stage~\cite{shao2023superglobal}, which introduced a \textit{global-to-global} system.
Our work goes beyond these existing directions to introduce the \textit{local-to-global} paradigm, which presents significant advantages compared to previous ones.
We build on top of recently-proposed techniques for efficient local feature search and global feature re-ranking, which enables localized search at large scale and effective re-ranking that merges information across the query and shortlisted images.

\noindent\textbf{Global feature-based re-ranking} is a recent idea to improve retrieval systems, as introduced by~\cite{shao2023superglobal}, to efficiently reorder the shortlisted images found in the initial search stage.
Its key insight is to leverage pairwise similarities among all shortlisted images, which can guide an aggregation process that refines the global features for re-ranking purposes.
In this work, we go beyond to introduce the usage of multidimensional scaling~\cite{nasir2018mds} to create a new global feature at re-ranking time.
This can help leverage detailed pairwise local feature-based similarities between the images, by converting them on-the-fly into an embedding space which respects those similarities.
Such a re-ranking embedding can then be used in the procedure introduced in~\cite{shao2023superglobal}, to refine all of the embeddings based on pairwise similarities, enabling a significantly improved final list of shortlist images.
\section{Local-to-Global Image Retrieval}
\label{sec:method}

\begin{figure*}[t]
    \centering
    \includegraphics[width=\textwidth]{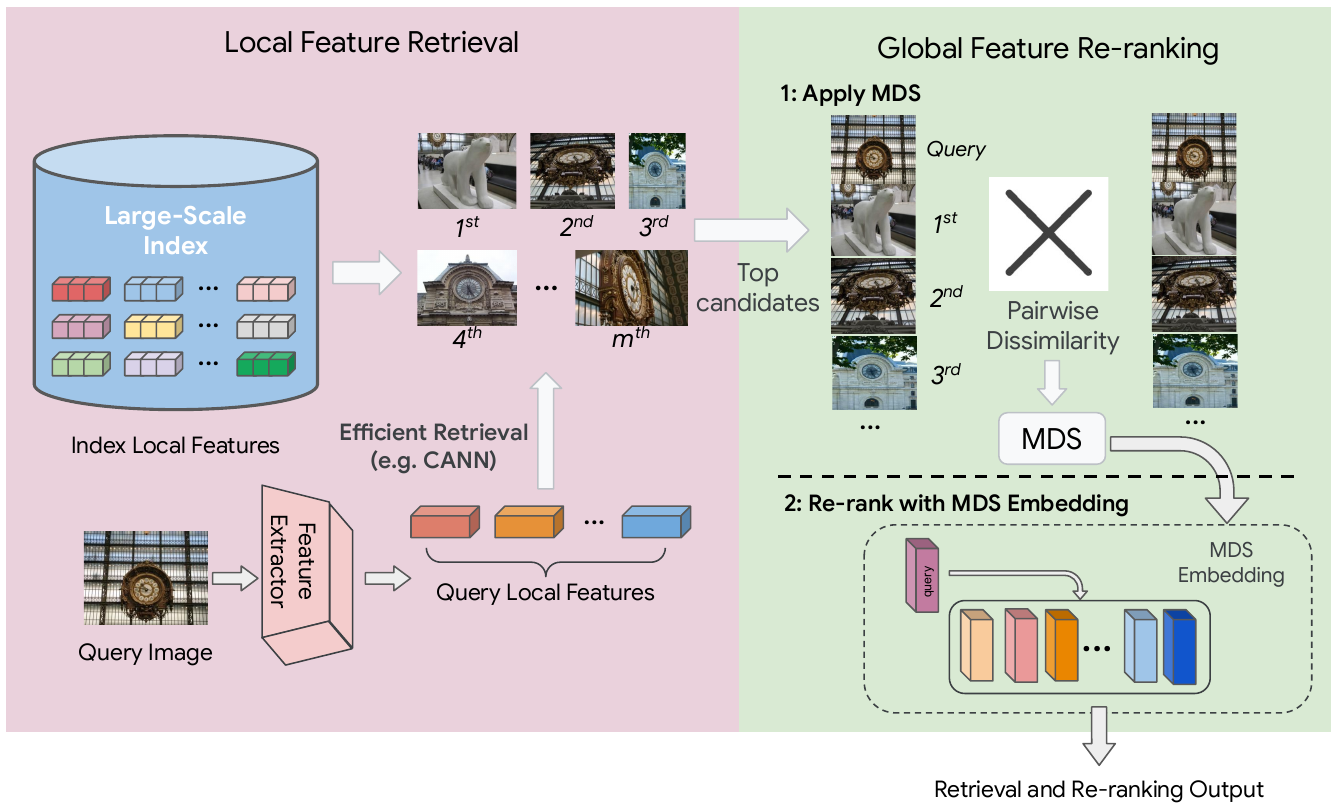}
    \caption{\textbf{Block diagram of the proposed Local-to-Global (L2G) retrieval system.}
    Left: given a query image, we extract local features which are used for efficient search via CANN~\cite{aiger2023cann}.
    Right: the top-ranked candidates from the local feature search stage undergo a re-ranking process leveraging embeddings computed on-the-fly via multidimensional scaling (MDS), based on the pairwise dissimilarities of the query and shortlisted database images.
    % Our L2G diagram consists of efficient local feature retrieval (left panel) and global feature reranking (right panel), in which reranking is conducted through MDS embeddings build upon the pairwise (dis)similarities. %\andre{TODO: Bingyi and/or Kaifeng to draw figure with system's block diagram.}
    }
    \label{fig:method}
    \vspace{-10pt}
\end{figure*}

The initial approaches to image retrieval relied heavily on hand-crafted local features. However, extracting and matching a large number of these features can be computationally expensive. With the advent of deep learning, global features, which capture the overall content and semantics of an image in a single vector representation, gained prominence.
While global features offer efficiency, local features provide finer granularity and robustness to changes in viewpoint or occlusions.  Some methods try to combine the strengths of both. For example, they might use global features for an initial retrieval and then re-rank results using local feature matching.

Recent advances in efficient algorithms have enabled the use of local features even in the initial retrieval stage, not just as a post-processing step. 
%This approach, which we call Local-to-Global Image Retrieval, leverages the complementary strengths of local and global features for improved accuracy.
One such approach is Constrained Approximate Nearest Neighbors (CANN)~\cite{aiger2023cann}, initially proposed for visual localization. CANN employs a novel nearest neighbor search strategy that efficiently finds the best matches in both appearance and geometry space using only local features and (asymmetric) Chamfer similarity.  As a byproduct, the authors demonstrated the potential of CANN for efficient image retrieval using local features in the first stage. They showed that a simple weighted average of rankings obtained from both global and local features significantly improves retrieval quality.
Another approach, MUVERA~\cite{dhulipala2023muvera}, utilizes multiple vector embeddings. It could also be used for efficient local feature-based image retrieval using the Chamfer similarity.

In this work, we present a novel and more effective method for merging local and global features, illustrated in Fig.~\ref{fig:method}.
Leveraging local features for the initial search stage and global features for re-ranking, our Local-to-Global (L2G) method delivers effective retrieval performance. 
We also introduce a natural way to integrate the re-ranking technique proposed by Shao et al.~\cite{shao2023superglobal} with any (dis)similarity, not necessarily a metric. 

\subsection{Re-ranking Global Features} 
\label{sec:re-ranking}
Shao et al.~\cite{shao2023superglobal} introduced a novel image re-ranking method that refines global features for improved retrieval performance. This method, designed to be plugged into any existing retrieval system, operates solely on global features, offering a significant efficiency advantage over conventional re-ranking techniques that rely on computationally expensive local features. Notably, it was the first solution to address both retrieval and re-ranking using only global image features.

Our work builds upon this concept but deviates from the convention of relying solely on global features. We leverage recent advances in efficient local feature-based retrieval, specifically CANN~\cite{aiger2023cann}, to incorporate local information into the re-ranking process.  To achieve this, we convert the ranking produced by CANN, which is based on non-metric similarity, into points in an embedding space. These points can then be treated as ``global features" representing the local feature information.
This transformation allows us to seamlessly integrate local and global features of any similarity. By merging these ``globalized" local features with existing global features, we can effectively utilize the re-ranking method of Shao et al.~\cite{shao2023superglobal}, resulting in improved retrieval accuracy.

\subsection{Enhanced Re-ranking with MDS}

Consider the following well known problem: Given pairwise dissimilarities, reconstruct a set of points that preserves pairwise distances. 
Multidimensional scaling (MDS)~\cite{nasir2018mds} is a family of widely used techniques for mapping data from a high-dimensional to a lower-dimensional space and for visualizing data. 
It is also a known method to reconstruct a set of points in high dimensional space from their pairwise distances. 
Given a set of dissimilarities, one can ask whether these values are distances and, moreover, whether they can even be interpreted as Euclidean distances. 
Given a dissimilarity (distance) matrix $D = (d_{ij})$, MDS seeks to find $x_1, \ldots, x_n \in R^p$ such that $d_{ij} \approx ||x_i-x_j||$ as close as possible. 
For certain cases, for some large $p$, there exists a configuration $x_1, \ldots, x_n$ with exact distance match $d_{ij} = ||x_i-x_j||$. 

In such a case the distance involved is called Euclidean.
There are, however, cases where the dissimilarity is distance, but there exists no configuration in any $p$ with perfect match $d_{ij} = ||x_i-x_j||$, for some $i, j$. 
Such a distance is called non-Euclidean.
Classical MDS is the case where we have Euclidean distance matrix $D = (d_{ij})$. 
In this case, we have a globally optimal solution (non-unique since any rigid transformation of it is always a solution) at some dimension $p$ and there are efficient methods to find it. 
Metric MDS is where we are given a dimension $p$ and a monotone function $f$, and we seek to find an optimal configuration $X \subset R^p$ that gives $f (d_{ij}) \approx \hat{d}_{ij} = ||x_i-x_j||_2$ as close as possible. 
In many applications of MDS, dissimilarities are known only by their rank order, and the spacing between successively ranked dissimilarities is of no interest or is unavailable. 
This is the Non Metric MDS where we are given a dimension $p$, and we seek to find an optimal configuration $X \subset R^p$ that gives $f (d_{ij}) \approx d^*_{ij} = ||x_i-x_j||_2$ as close as possible. Different approaches exist for Non-Metric MDS, including stress minimization techniques like SMACOF~\cite{deleeuw1977applications}, which minimize a stress function that quantifies the discrepancy between the disparities and the distances in the embedded space. Unlike metric MDS, here $f$ is much general and only implicitly defined. $f(d_{ij})=d^*_{ij}$ are called disparities, which only preserve the order of $d_{ij}$,

\[
d_{ij} < d_{kl} \iff f(d_{ij}) \leq f(d_{kl}) \iff d^*_{ij} \leq d^*_{kl}
\]
% \begin{equation}
% d_{ij} < d_{kl} \iff f(d_{ij}) \leq f(d_{kl}) \iff d^*_{ij} \leq d^*_{kl}
% \end{equation}

Our goal is to find an embedding in a high-dimensional space for the dissimilarity matrix of index and query images.
%, enabling the use of the re-ranking method proposed by Shao et al.~\cite{shao2023superglobal}. 
This method requires a complete set of pairwise distances between all images (both index and query) to perform nearest neighbor operations (for averaging) within a metric space.

However, our data presents two key challenges:
(1) 
Non-metric dissimilarities: The initial ranking we obtain from local feature matching is non-metric. This means it may not satisfy the properties of a distance function, such as the triangle inequality. Directly applying the re-ranking method, which assumes a metric space, would lead to inconsistencies.
(2) 
 Incomplete distances signify that the relationships between certain pairs of data points are unknown. This absence of information can lead to a distorted representation of the data's overall structure in the embedding. In our case, obtaining pairwise distances between all pairs of index images is impractical, as it would require computation that grows quadratically with the dataset size. To maintain efficiency and scalability, we instead utilize a sparse distance matrix. In this sparse representation, each image only stores distances to its nearest neighbors, determined using an existing retrieval system.

To address these challenges, we turn to Multidimensional Scaling (MDS). While various MDS methods exist, we found the classic landmark MDS~\cite{ST2004,venna2010global}, which relies on eigenvalue decomposition and Nyström approximation, less effective in our case. This is likely because landmark MDS generally requires metric distances and struggles with the non-metric nature of our dissimilarities.

Instead, we employ SMACOF (Scaling by MAjorizing a COmplicated Function), an iterative optimization algorithm first introduced in~\cite{deleeuw1977applications}. SMACOF minimizes a stress function that quantifies the discrepancy between the given dissimilarities and the distances in the reconstructed configuration. This iterative approach is well-suited for handling non-metric dissimilarities and incomplete data.

Specifically, SMACOF allows us to: (i) Handle non-metric dissimilarities: Effectively address the non-metric nature of our ranking data.
(ii) Complete the dissimilarity matrix: Infer the missing entries to construct a complete distance matrix required for the re-ranking method.
(iii) Weight dissimilarities: Assign weights to different dissimilarities based on their reliability and importance, potentially improving the embedding quality.
%This revised version clarifies the challenges posed by your data and explains why you chose SMACOF over other MDS methods. It also highlights the specific benefits of SMACOF in addressing those challenges.

\paragraph{Complexity} The computational complexity of MDS varies significantly depending on the specific method and the size of the data. Here's a breakdown of the complexity for some common MDS approaches: Classical MDS  (Eigenvector-based) has complexity of $O(N^3)$ where $N$ is the number of data points. This is dominated by the eigenvalue decomposition of the $N \times N$ dissimilarity matrix. SMACOF's complexity is $O(N^2)$ per iteration and the number of iterations in our case is fairly small ($\approx$ 5-10). Landmark MDS complexity can be significantly lower than classical MDS for large datasets. By using a smaller set of ``landmark" points (say, $L$ landmarks), the complexity can be reduced to approximately $O(NL^2 + L^3)$. This makes it more scalable for large $N$ when $L << N$. FastMap~\cite{faloutsos1995fastmap} complexity is $O(Nk)$ where $N$ is the number of data points and $k$ is the desired dimensionality of the embedding.

\subsection{On-the-fly MDS per query}

Similarity embedding using MDS can be viewed as generating ``global features" derived from a specific similarity measure, as opposed to generic embeddings trained on separate data. We take a similarity matrix and compute a global representation of the points that best preserves these similarities.

We assume that for an appropriate similarity (e.g., Chamfer), we have an efficient method (CANN, MUVERA) to index a large dataset and retrieve the top-$k$ most similar images for any given query. We denote this method as EFF-INDEX-QUERY. Using EFF-INDEX-QUERY, we can pre-compute the top-k nearest index images for each image in the index offline. This results in a sparse set of pairwise distances, denoted as INDEX-SPARSE-DISTANCES, with a size linear in the number of data points since $k$ is constant.

One approach is to apply MDS to the entire index at indexing time, creating and storing the embeddings just as we would store global features. This can be achieved in approximately linear time using fast MDS methods. At query time, we obtain the distances from the query to the top index images using EFF-INDEX-QUERY and then compute the query image embedding using the same methods used in landmark MDS, which requires only a constant number of images. With all embeddings computed, we can proceed with re-ranking in a metric space.

However, we propose a more efficient alternative that avoids applying MDS to the entire index. Instead, we compute MDS specifically for each query image and its top-$k$ ranked index images, retrieved initially using EFF-INDEX-QUERY. This localized approach generates an embedding for only $k+1$ points, significantly reducing the computational burden.

For each query image, we use EFF-INDEX-QUERY to retrieve the top-ranked images and obtain their pairwise distances from INDEX-SPARSE-DISTANCES. If the distance between a pair of images is not present in INDEX-SPARSE-DISTANCES, we set it to $1$ (the maximum possible distance). 
We then apply standard MDS to the $(k+1) \times (k+1)$ similarity matrix to obtain an Euclidean embedding of the query and its top-$k$ neighbors. This computation takes $O(k^2)$ time, where $k$ is typically a constant. While this can be further improved to $O(k)$ using landmark MDS or other fast approximate MDS methods, we opted for standard MDS in our experiments due to its efficiency for moderate $k$ and our focus on demonstrating the core concept.

This localized embedding strategy allows us to apply efficient re-ranking techniques, similar to those in~\cite{shao2023superglobal}, without requiring index-time embedding. While the embedding is recomputed for each query, the overall computational cost remains manageable due to the small number of points involved.

\section{Experiments}
\label{sec:experiments}

\subsection{Experimental Setup}
Our experiments are conducted on well-established benchmarks. Concretely, we use Oxford \cite{Philbin07} and Paris \cite{Philbin2008} with revisited annotations, referred to as $\mathcal{R}$Oxf  and $\mathcal{R}$Par, respectively. There are 4993 (6322) database images in the $\mathcal{R}$Oxf ($\mathcal{R}$Par) dataset, and each dataset contains a query set with 70 images. Large-scale results are further reported with the $\mathcal{R}$1M distractor set \cite{radenovic2018revisiting}, which contains 1M database images. 

We report the effectiveness of SMACOF MDS~\cite{deleeuw1977applications} re-ranking when combined with local feature retrieval. 
We leverage the FIRE \cite{weinzaepfel2022learning} image features, and to ensure high retrieval efficiency, we follow the algorithm proposed by the CANN paper \cite{aiger2023cann} and utilize the official implementation\footnote{\url{https://github.com/google-research/google-research/tree/master/cann}} and tune it on the $\mathcal{R}$Oxford dataset. For re-ranking with global features, we employ a weighted average between the MDS embeddings and SuperGlobal global features, where the MDS embeddings are obtained directly using the pairwise FIRE Chamfer similarities. We further tune the following hyperparameters on the $\mathcal{R}$Oxford dataset.

- $\epsilon$ which controls the convergence threshold for the MDS algorithm.

- $p$ (power), the modulation parameter that adjusts the influence of small and large distances in the Chamfer similarity metric.

- $w$ (weight) that determines the relative importance of the SuperGlobal features and the MDS embeddings when combining them for re-ranking.

- $k$ (top ranked for MDS) that specifies the number of top-ranked images from the initial retrieval that are used for MDS embedding. It’s important to note that this is distinct from the M parameter used in the re-ranking stage after we have the embedding, which determines the number of top-ranked images considered for neighborhood analysis. In our experiments, embedding re-ranking (M) is always conducted among the top 1600 candidates. 

All hyperparameters were tuned on a small sample (1000 images) of Oxford only. The optimal parameters found experimentally are $w=0.19,p=0.01,\epsilon=0.1,k=700$. We also set different SG re-ranking hyper parameters: $M=1600,k=6,\beta=0.31$ (see~\cite{shao2023superglobal}). All hyper parameters are the same for all the experiments. We believe the complementary nature of these embeddings is key to the performance improvements. 
We use the standard mean Average Precision (mAP) as the evaluation metric.

\begin{table*}[t]
\resizebox{\textwidth}{!}{
\begin{tabular}{lcccc|cccc}
\setlength\tabcolsep{3pt}
& \multicolumn{4}{c|}{Medium} & \multicolumn{4}{c}{Hard} \\
Method & $\mathcal{R}$Oxf & $\mathcal{R}$Oxf+1M & $\mathcal{R}$Par & $\mathcal{R}$Par+1M & $\mathcal{R}$Oxf & $\mathcal{R}$Oxf+1M & $\mathcal{R}$Par & $\mathcal{R}$Par+1M \\
\hline
(1) Global feature retrieval \\
RN50-DELG~\cite{cao2020delg} & 73.6 & 60.6 & 85.7 & 68.6 & 51.0 & 32.7 & 71.5 & 44.4 \\
RN101-DELG~\cite{cao2020delg} & 76.3 & 63.7 & 86.6 & 70.6 & 55.6 & 37.5 & 72.4 & 46.9 \\
RN50-DOLG~\cite{yang2021dolg} & 80.5 & 76.6 & 89.8 & 80.8 & 58.8 & 52.2 & 77.7 & 62.8 \\
RN101-DOLG~\cite{yang2021dolg} & 81.5 & 77.4 & 91.0 & 83.3 & 61.1 & 54.8 & 80.3 & 66.7 \\
RN50-CVNet~\cite{lee2022cvnet} & 81.0 & 72.6 & 88.8 & 79.0 & 62.1 & 50.2 & 76.5 & 60.2 \\
RN101-CVNet~\cite{lee2022cvnet} & 80.2 & 74.0 & 90.3 & 80.6 & 63.1 & 53.7 & 79.1 & 62.2 \\
RN50-SuperGlobal (No re-ranking)~\cite{shao2023superglobal} & 83.9 & 74.7 & 90.5 & 81.3 & 67.7 & 53.6 & 80.3 & 65.2 \\
RN101-SuperGlobal (No re-ranking)~\cite{shao2023superglobal} & 85.3 & 78.8 & 92.1 & 83.9 & 72.1 & 61.9 & 83.5 & 69.1 \\
\hline
(2) Global feature retrieval + Local feature re-ranking \\
RN50-DELG (GV re-rank top 100)~\cite{cao2020delg} & 78.3 & 67.2 & 85.7 & 69.6 & 57.9 & 43.6 & 71.0 & 45.7 \\
RN101-DELG (GV re-rank top 100)~\cite{cao2020delg} & 81.2 & 69.1 & 87.2 & 71.5 & 64.0 & 47.5 & 72.8 & 48.7 \\
RN50-CVNet (Re-rank top 400)~\cite{lee2022cvnet} & 87.9 & 80.7 & 90.5 & 82.4 & 75.6 & 65.1 & 80.2 & 67.3 \\
RN101-CVNet (Re-rank top 400)~\cite{lee2022cvnet} & 87.2 & 81.9 & 91.2 & 83.8 & 75.9 & 67.4 & 81.1 & 69.3 \\
\hline
(3) SuperGlobal retrieval + Re-ranking \\
RN50-SuperGlobal (Re-rank top 400)~\cite{shao2023superglobal} & 88.8 & 80.0 & 92.0 & 83.4 & 77.1 & 64.2 & 84.4 & 68.7 \\
RN101-SuperGlobal (Re-rank top 400)~\cite{shao2023superglobal} & 90.9 & 84.4 & 93.3 & 84.9 & 80.2 & 71.1 & 86.7 & 71.4 \\
AMES (600,600) (Re-rank top 1600)~\cite{suma2024ames} &  \bf{93.6} & \underline{88.2} & \underline{95.3} & \underline{90.1} & \bf{84.8} &  \underline{77.7} & \underline{90.7} & \underline{82.0} \\
\hline
(4) Local feature retrieval + Global feature re-ranking \\
L2G CANN-FIRE + MDS re-ranking (\textbf{Ours}) & \underline{92.9} & \bf{90.5} & \bf{97.1} & \bf{92.1} & \underline{83.0} & \bf{79.8} & \bf{91.7} & \bf{83.4} \\
\hline
\end{tabular}}
\caption{\label{tab:oxford_paris}\textbf{Comparison to the state of the art.} Results of our L2G approach, compared to state-of-the-art methods on $\mathcal{R}$Oxford and $\mathcal{R}$Paris, on their base and extended ``+1M" versions. Best results per dataset in \textbf{bold}, second-best \underline{underlined}.}
\end{table*}

% By placing the figure* definition early, we give LaTeX a better chance to fit it on an earlier page.
% The [htbp] specifier allows placement here, top, bottom, or on a float page (made possible by stfloats).
\begin{figure*}[htbp]
    \centering
    \includegraphics[width=0.99\textwidth]{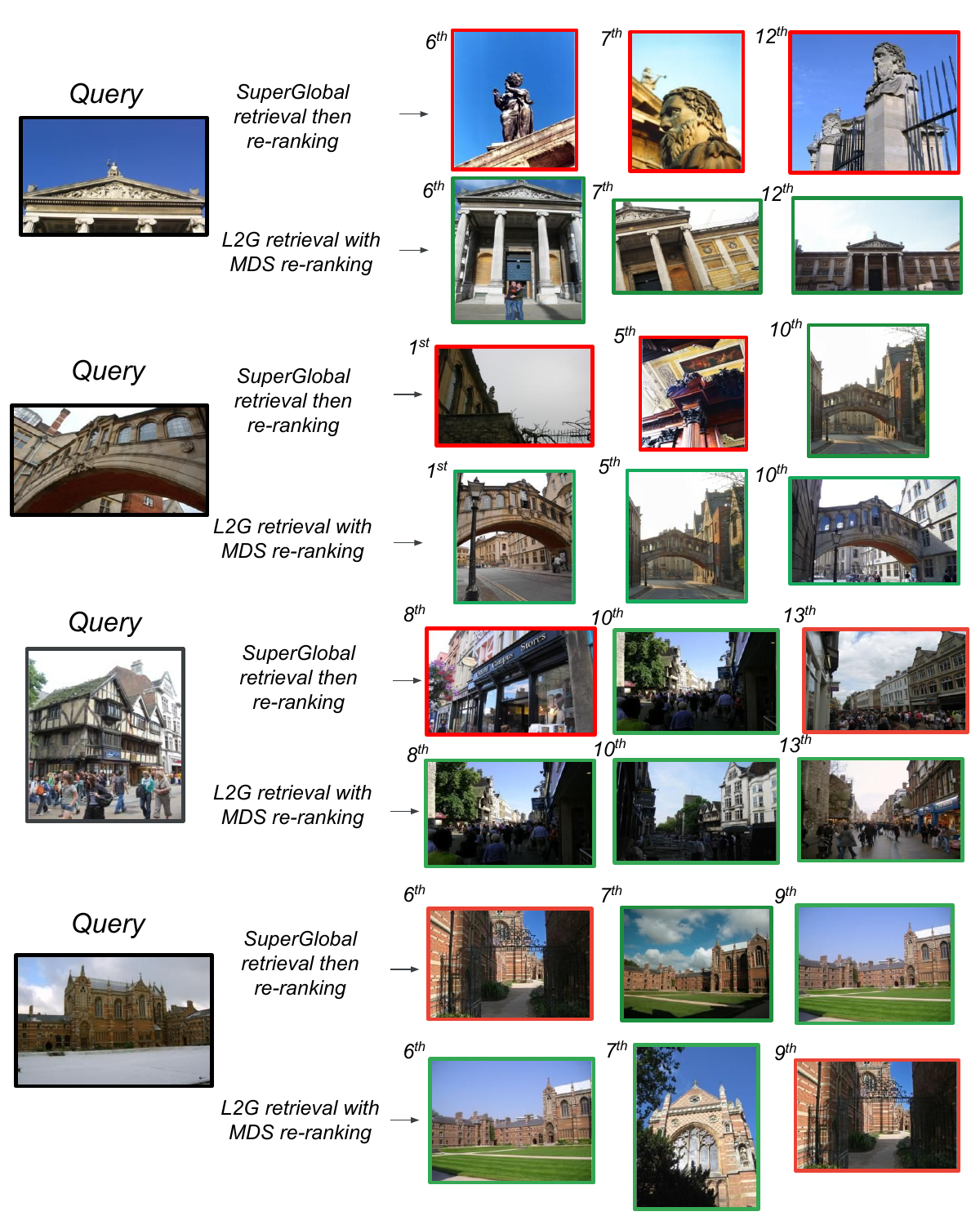}
    \caption{\textbf{Qualitative results.}
    Examples comparing our L2G method with FIRE~\cite{weinzaepfel2022learning} local features against SuperGlobal~\cite{shao2023superglobal} retrieval and re-ranking, on representative queries from the $\mathcal{R}$Oxf and $\mathcal{R}$Par datasets.}
    \label{fig:qualitative}
\end{figure*}

\subsection{Results}
We compare our results with state-of-the-art models in Table \ref{tab:oxford_paris}. The results are split into four settings: (1) Global feature retrieval; (2) Global feature retrieval + Local feature re-ranking (G2L); (3) SuperGlobal retrieval + Re-ranking; and (4) Local feature retrieval + Global feature re-ranking (L2G).

Our L2G approach, though applied to the FIRE feature from 2022, achieves state-of-the-art performance, exceeding modern approaches like SuperGlobal (2023) and AMES (2024). In particular, our method is extremely effective in large-scale settings, achieving $79.8\%$ in $\mathcal{R}$Oxf+1M Hard and $83.4\%$ in $\mathcal{R}$Par+1M Hard, beating the best AMES results by $2.1\%$ and $1.4\%$, respectively.

\begin{figure}[t]
    \centering
    \includegraphics[width=\columnwidth]{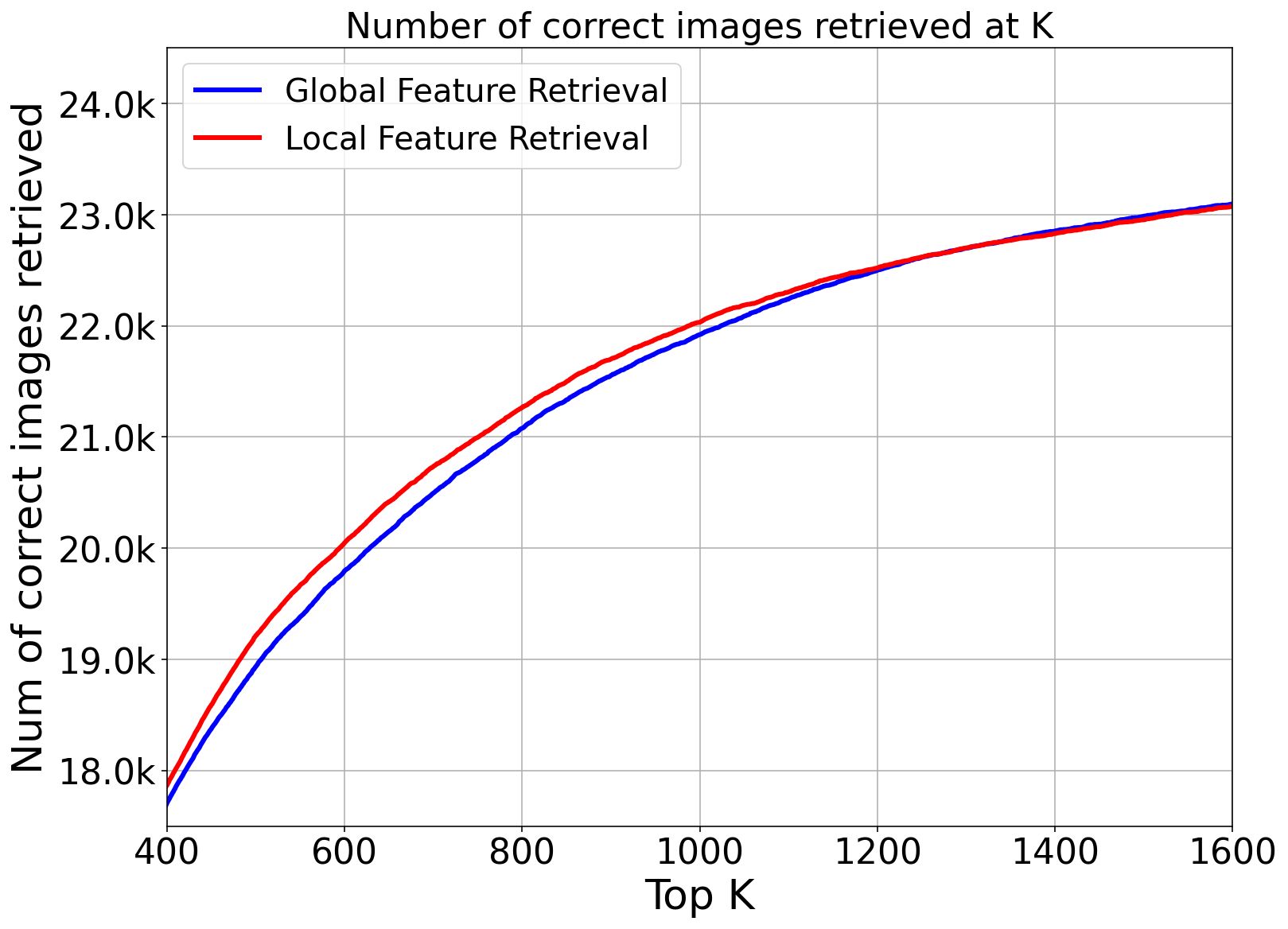}
    \caption{Number of correct images retrieved at top $K$ via local and global feature retrieval on the $\mathcal{R}$Paris dataset.}
    \label{fig:recall}
\end{figure}

We conduct a study on $\mathcal{R}$Paris (Fig. \ref{fig:recall}) and find that for the top 1600 candidates, local and global retrieval find about the same number of correct images. This reveals that our MDS re-ranking is critical, as it significantly outperforms other methods on the same set of candidates, showing the importance of using an embedding sensitive to localized similarities.

In Fig. \ref{fig:qualitative}, we provide qualitative results comparing our L2G approach with SuperGlobal. Our L2G method retrieves matching images (green boxes) where SuperGlobal fails (red boxes), resolving many common failure cases.

\subsection{Ablation Study}
We conducted an ablation study to analyze the impact of different components on our method's performance (Table~\ref{tab:ablation}).
\textbf{Without SuperGlobal re-ranking process}, performance drops, showing the value of the refinement update. \textbf{Without merging final features with SuperGlobal} ($w=1.0$) also hurts performance, confirming the synergy between the two feature types. When we \textbf{Replace MDS re-ranking by SuperGlobal re-ranking}, the poor results show the incompatibility of using a non-metric similarity in a metric-space re-ranker. Finally, when we \textbf{Replace FIRE local similarity by AMES}, our framework achieves the best overall performance, demonstrating its power and modularity.

\begin{table*}[t]
\centering
\begin{tabular}{|l|c|c|}
\hline
\textbf{Configuration} & \textbf{$\mathcal{R}$Paris} & \textbf{$\mathcal{R}$Oxford}\\
\hline
Full Model & 91.7 & 83.0  \\
\hline
- Without SuperGlobal re-ranking process & 84.6 & 73.3  \\
- Without merging final similarities with SuperGlobal & 89.4 & 81.6  \\
- Replace MDS re-ranking by SuperGlobal re-ranking & 86.4 & 72.4 \\
- Replace FIRE local similarity by AMES & 93.6 & 85.2 \\
\hline
\end{tabular}
\caption{\label{tab:ablation}Ablation Study: mAP for $\mathcal{R}$Oxford and $\mathcal{R}$Paris Hard with various components of the pipeline. See the text for details.}
\end{table*}

\subsection{Computational Cost and Memory}

\noindent
\textbf{Computational Cost.} Our full L2G pipeline has a query time of approximately \textbf{0.7 seconds} on the $\mathcal{R}$Oxf+1M dataset. This is broken down into $\approx$0.2s for the initial local feature retrieval using CANN and $\approx$0.5s for the subsequent on-the-fly MDS embedding computation and re-ranking of the top 1600 candidates. These timings were measured on a 24-core CPU.

\noindent
\textbf{Memory Requirements.} In terms of memory, our approach requires storing local features for the database images. For the $\mathcal{R}$Oxf+1M dataset, using 600 FIRE features per image without compression results in a memory footprint of approximately \textbf{21kB per image}. This is higher than typical global feature methods or AMES with PQ8 compression (~10kB per image), but we argue this is justified by the significant accuracy gains.
\section{Conclusions}

Our work presents a new \textit{local-to-global} image retrieval system, leveraging local image features for the initial large-scale search and global image features, induced by the MDS of the local similarity, for the re-ranking stage.
This is a significant departure from today's conventional \textit{global-to-local} paradigm, helping overcome issues with partial matches at large scale and insufficient local information for re-ranking a short list of images.
Notably, we introduce a novel global feature re-ranking process which can effectively leverage local similarities by converting these similarities into a new embedding space which respects those.
Leveraging multidimensional scaling, these re-ranking embeddings significantly boost performance.
Our experiments showcase strong results in conventional image retrieval datasets.

\noindent\textbf{Future work.}
This work with MDS, particularly its fast variants, opens exciting possibilities for using it with diverse similarity measures, including learned ones like those in AMES~\cite{suma2024ames}. The key observation that embedding from pairwise distances requires only a constant number of pairs suggests a novel and efficient approach to building indexing and query systems for any generic similarity. This research direction could lead to significant advancements in similarity search, enabling more efficient and accurate retrieval across diverse domains and applications.

\noindent\textbf{Limitations.}
The local feature retrieval process is more expensive than the global feature one, however the use of CANN makes it very efficient and competitive, while at the same time providing better results.
Our re-ranking technique is more expensive than SuperGlobal as it requires the multidimensional scaling step to compute the re-ranking embeddings -- however, this can be efficiently computed, as previously discussed, while at the same time enhancing the accuracy of the system.

%%%%%%%%% REFERENCES
\clearpage
\small
\bibliographystyle{ieee_fullname}
\bibliography{egbib}

\begin{thebibliography}{10}\itemsep=-1pt

\bibitem{aiger2023cann}
D. Aiger, A. Araujo, and S. Lynen.
\newblock {Yes, we CANN: Constrained Approximate Nearest Neighbors for Local
  Feature-Based Visual Localization}.
\newblock In {\em Proc. ICCV}, 2023.

\bibitem{babenko2014neural}
A. Babenko, A. Slesarev, A. Chigorin, and V. Lempitsky.
\newblock {Neural Codes for Image Retrieval}.
\newblock In {\em Proc. ECCV}, 2014.

\bibitem{bay2008speeded}
H. Bay, A. Ess, T. Tuytelaars, and L. Van~Gool.
\newblock {Speeded-Up Robust Features (SURF)}.
\newblock {\em CVIU}, 2008.

\bibitem{cao2020delg}
B. Cao, A. Araujo, and J. Sim.
\newblock {Unifying Deep Local and Global Features for Image Search}.
\newblock In {\em Proc. ECCV}, 2020.

\bibitem{chen2023infoseek}
Y. Chen, H. Hu, Y. Luan, H. Sun, S. Changpinyo, A. Ritter, and M.-W. Chang.
\newblock {Can Pre-trained Vision and Language Models Answer Visual
  Information-Seeking Questions?}
\newblock In {\em Proc. EMNLP}, 2023.

\bibitem{deleeuw1977applications}
Jan De~Leeuw.
\newblock Applications of convex analysis to multidimensional scaling.
\newblock In J.~R. Barra, F. Brodeau, G. Romier, and B. Van~Cutsem, editors,
  {\em Recent Developments in Statistics}, pages 133--145. North-Holland
  Publishing Company, 1977.

\bibitem{dhulipala2023muvera}
Laxman Dhulipala, Majid Hadian, Rajesh Jayaram, Jason Lee, and Vahab Mirrokni.
\newblock Muvera: Multi-vector retrieval via fixed dimensional encodings.
\newblock In {\em Advances in Neural Information Processing Systems}, 2023.

\bibitem{faloutsos1995fastmap}
Christos Faloutsos and King-Ip Lin.
\newblock Fastmap: A fast algorithm for indexing, data-mining and visualization
  of traditional and multimedia datasets.
\newblock In {\em Proceedings of the 1995 ACM SIGMOD international conference
  on Management of data}, pages 163--174, 1995.

\bibitem{gordo2017end}
A. Gordo, J. Almazan, J. Revaud, and D. Larlus.
\newblock {End-to-end Learning of Deep Visual Representations for Image
  Retrieval}.
\newblock {\em IJCV}, 2017.

\bibitem{hu2023reveal}
Z. Hu, A. Iscen, C. Sun, Z. Wang, K.-W. Chang, Y. Sun, C. Schmid, D. Ross, and
  A. Fathi.
\newblock {REVEAL: Retrieval-Augmented Visual-Language Pre-Training with
  Multi-Source Multimodal Knowledge Memory}.
\newblock In {\em Proc. CVPR}, 2023.

\bibitem{jegou2012aggregating}
H. J\'{e}gou, F. Perronnin, M. Douze, J. Sanchez, P. Perez, and C. Schmid.
\newblock {Aggregating Local Image Descriptors into Compact Codes}.
\newblock {\em PAMI}, 2012.

\bibitem{krause20133d}
J. Krause, M. Stark, J. Deng, and L. Fei-Fei.
\newblock {3D Object Representations for Fine-Grained Categorization}.
\newblock In {\em Proc. ICCV Workshops}, 2013.

\bibitem{lee2022cvnet}
S. Lee, H. Seong, S. Lee, and E. Kim.
\newblock {Correlation Verification for Image Retrieval}.
\newblock In {\em Proc. CVPR}, 2022.

\bibitem{liu2016cvpr}
Z. Liu, P. Luo, S. Qiu, X. Wang, and X. Tang.
\newblock {Deepfashion: Powering Robust Clothes Recognition and Retrieval with
  Rich Annotations}.
\newblock In {\em Proc. CVPR}, 2016.

\bibitem{Lowe2004}
D. Lowe.
\newblock {Distinctive Image Features from Scale-Invariant Keypoints}.
\newblock {\em IJCV}, 2004.

\bibitem{mensink2023encvqa}
T. Mensink, J. Uijlings, L. Castrejon, A. Goel, F. Cadar, H. Zhou, F. Sha, A.
  Araujo, and V. Ferrari.
\newblock {Encyclopedic VQA: Visual Questions About Detailed Properties of
  Fine-Grained Categories}.
\newblock In {\em Proc. ICCV}, 2023.

\bibitem{Ng2020SOLARSL}
T. Ng, V. Balntas, Y. Tian, and K. Mikolajczyk.
\newblock {SOLAR: Second-Order Loss and Attention for Image Retrieval}.
\newblock In {\em Proc. ECCV}, 2020.

\bibitem{Nister2006vocabulary}
D. Nist{\'e}r and H. Stewenius.
\newblock {Scalable Recognition with a Vocabulary Tree}.
\newblock In {\em Proc. CVPR}, 2006.

\bibitem{noh2017large}
H. Noh, A. Araujo, J. Sim, T. Weyand, and B. Han.
\newblock {Large-Scale Image Retrieval with Attentive Deep Local Features}.
\newblock In {\em Proc. ICCV}, 2017.

\bibitem{peng2021rp2k}
J. Peng, C. Xiao, and Y. Li.
\newblock {RP2K: A Large-Scale Retail Product Dataset for Fine-Grained Image
  Classification}.
\newblock {\em arXiv:2006.12634}, 2021.

\bibitem{Philbin07}
J. Philbin, O. Chum, M. Isard, J. Sivic, and A. Zisserman.
\newblock {Object Retrieval with Large Vocabularies and Fast Spatial Matching}.
\newblock In {\em Proc. CVPR}, 2007.

\bibitem{Philbin2008}
J. Philbin, O. Chum, M. Isard, J. Sivic, and A. Zisserman.
\newblock {Lost in Quantization: Improving Particular Object Retrieval in Large
  Scale Image Databases}.
\newblock In {\em Proc. CVPR}, 2008.

\bibitem{radenovic2018revisiting}
F. Radenovi{\'c}, A. Iscen, G. Tolias, Y. Avrithis, and O. Chum.
\newblock {Revisiting Oxford and Paris: Large-Scale Image Retrieval
  Benchmarking}.
\newblock In {\em Proc. CVPR}, 2018.

\bibitem{Revaud2019ICCV}
J. Revaud, J. Almazan, R.~S. Rezende, and C.~R. Souza.
\newblock {Learning With Average Precision: Training Image Retrieval With a
  Listwise Loss}.
\newblock In {\em Proc. ICCV}, October 2019.

\bibitem{nasir2018mds}
N. Saeed, H. Nam, M. Haq, and D. Saqib.
\newblock {A Survey on Multidimensional Scaling}.
\newblock {\em ACM Comput. Surv.}, 2018.

\bibitem{shao2023superglobal}
S. Shao, K. Chen, A. Karpur, Q. Cui, A. Araujo, and B. Cao.
\newblock {Global Features are All You Need for Image Retrieval and Reranking}.
\newblock 2023.

\bibitem{ST2004}
V.~d. Silva and J.~B. Tenenbaum.
\newblock Sparse multidimensional scaling using landmark points.
\newblock {\em Technical Report (Stanford University)}, 2004.

\bibitem{Sivic2003}
J. Sivic and A. Zisserman.
\newblock {Video Google: A Text Retrieval Approach to Object Matching in
  Videos}.
\newblock In {\em ICCV}, 2003.

\bibitem{song2016deep}
H. Song, Y. Xiang, S. Jegelka, and S. Savarese.
\newblock {Deep Metric Learning via Lifted Structured Feature Embedding}.
\newblock In {\em Proc. CVPR}, 2016.

\bibitem{suma2024ames}
P. Suma, G. Kordopatis-Zilos, A. Iscen, and G. Tolias.
\newblock {AMES: Asymmetric and Memory-Efficient Similarity Estimation for
  Instance-level Retrieval}.
\newblock In {\em Proc. ECCV}, 2024.

\bibitem{tan2021rrt}
F. Tan, J. Yuan, and V. Ordonez.
\newblock {Instance-level Image Retrieval using Reranking Transformers}.
\newblock In {\em Proc. ICCV}, 2021.

\bibitem{teichmann2019detect}
M. Teichmann, A. Araujo, M. Zhu, and J. Sim.
\newblock {Detect-to-Retrieve: Efficient Regional Aggregation for Image
  Search}.
\newblock In {\em CVPR}, 2019.

\bibitem{tolias2015image}
G. Tolias, Y. Avrithis, and H. Jegou.
\newblock {Image Search with Selective Match Kernels: Aggregation Across Single
  and Multiple Images}.
\newblock {\em IJCV}, 2015.

\bibitem{tolias2020learning}
G. Tolias, T. Jenicek, and O. Chum.
\newblock {Learning and Aggregating Deep Local Descriptors for Instance-Level
  Recognition}.
\newblock In {\em ECCV}, 2020.

\bibitem{venna2010global}
Jarkko Venna, Jaakko Peltonen, Kristian Nybo, Helena Aidos, and Samuel Kaski.
\newblock Global versus local methods in nonlinear dimensionality reduction.
\newblock {\em Neural Networks}, 23(1):125--136, 2010.

\bibitem{weinzaepfel2022learning}
P. Weinzaepfel, T. Lucas, D. Larlus, and Y. Kalantidis.
\newblock {Learning Super-Features for Image Retrieval}.
\newblock In {\em ICLR}, 2022.

\bibitem{yang2021dolg}
M. Yang, D. He, M. Fan, B. Shi, X. Xue, F. Li, E. Ding, and J. Huang.
\newblock {DOLG: Single-Stage Image Retrieval with Deep Orthogonal Fusion of
  Local and Global Features}.
\newblock In {\em Proc. ICCV}, 2021.

\bibitem{ypsilantis2023uned}
N.-A. Ypsilantis, K. Chen, B. Cao, M. Lipovský, P. Dogan-Schonberger, G.
  Makosa, B. Bluntschli, M. Seyedhosseini, O. Chum, and A. Araujo.
\newblock {Towards Universal Image Embeddings: A Large-Scale Dataset and
  Challenge for Generic Image Representations}.
\newblock In {\em Proc. ICCV}, 2023.

\bibitem{ypsilantis2021met}
N.-A. Ypsilantis, N. Garcia, G. Han, S. Ibrahimi, N. Van~Noord, and G. Tolias.
\newblock {The Met Dataset: Instance-level Recognition for Artworks}.
\newblock In {\em Proc. NeurIPS Datasets and Benchmarks Track}, 2021.

\end{thebibliography}

\end{document}